# Near-Field Optical MIMO Communication with Polarization-dependent Metasurfaces

## Shamsi Soleimani [1], Kasra Rouhi [1] and Ali Momeni [2]

[1] School of Electrical Engineering, Iran University of Science and Technology, Tehran, Iran

[2] Laboratory of Wave Engineering, School of Electrical Engineering, Swiss Federal Institute of Technology in Lausanne (EPFL), Lausanne, Switzerland

## Abstract

The ability to control waves at the nanoscale has attracted considerable attention to ultrathin metasurface lenses (metalenses) in optical imaging and encryption systems. We propose an approach to active tuning metasurfaces by integrating an ultrathin layer of indium-tin-oxide (ITO) into a unit cell as an electro-optically tunable material. A proposed design features two orthogonal wings that can independently manipulate waves with corresponding orthogonal polarizations. The charge carrier concentration in the ITO accumulation layer is altered by modulating the applied bias voltage. This bias voltage generates phase variations at terahertz frequencies for the reflected transverse electric and transverse magnetic polarized waves. It is possible to move both focal points of a metalens without any physical movement by varying the bias voltage. In addition, this paper explores the application of virtually moving metalens for a novel multiple-input and multiple-output (MIMO) communication architecture. We demonstrate a communication system based on single-point binary data communication and hexadecimal orbital angular momentum (OAM) data communication. Then, orthogonal channels can be used for MIMO communication with high capacity. The proposed design paves the way for high-speed communications as well as polarization-controlled molecular imaging systems.

## I. Introduction

Metamaterials, artificially structured materials composed of subwavelength arrays of unit cells, can exhibit extraordinary properties beyond those accessible by natural materials. They were initially proposed for challenging fundamental laws and demonstrating negative refraction in the microwave regime. In subsequent research, metamaterials were used as versatile platforms to manipulate electromagnetic waves throughout the spectrum because of their extreme scalability. Metasurfaces are the equivalent version of two-dimensional metamaterials proposed as an effective method for arbitrary manipulation of amplitude, phase, and polarization of electromagnetic waves [1–5]. This unique feature of metasurfaces has recently attracted much attention due to its small size, the possibility of integration, and high flexibility in manipulating the wavefront. In recent years, tunable metasurfaces have attracted widespread attention because they provide significant opportunities for real-time wave manipulation. They can be used in various applications, including vortex beam generation [6], asymmetric transmission and wave manipulation [7,8], holography [9], wave-based analog computing [10,11], spatial wave control [12], and absorbers [13].

Terahertz lenses are widely used in spectrometers, terahertz communication systems, and millimeter and submillimeter imaging systems. Generally, there are three types of terahertz lenses: ordinary, Fresnel, and metasurface lenses [14]. In an ordinary terahertz lens, curved surfaces create a large space along with the thickness of the lens. Indeed, ordinary lenses concentrate the waves by accumulating phase differences along the path. The terahertz Fresnel lens is designed to reduce the thickness of ordinary lenses. On the other hand, in the metasurface lens (metalens), an extra phase adds to the wave, leading to constructive interference on the focal point. A metalens is a superficial lens that allows Fourier transform analysis. According to Fermat's principle, phase control can modify the wavefront [15,16]. Therefore, metasurface can create the required phase change and develop ultra-thin flat lenses with unique features and high performance. It also has high resolution and optimal performance for ordinary lenses.

In structures that are not sensitive to polarization, only a single function is considered for one or both polarizations. However, different behaviors can be extracted from orthogonal polarizations in an anisotropic unit cell design [17]. If the function of two radiations with $x$- and $y$-polarization can be separated, different reflections can be expected for each polarization, leading to independent beamforming. In [18], anisotropic transmissive metasurfaces are presented that enable simultaneous and independent control of amplitude and phase responses of two orthogonal polarizations. The transmission response of the suggested structure can give full phase coverage with widely adjustable amplitude and negligible cross-polarized components. In [19], the transmission response can be tuned to provide full phase coverage and minimal cross-polarized components. The designed metasurface is made up of two layers of graphene arrays that can be switched between two states by biasing the two graphene layers with the specified voltage and zero voltage, respectively. In this design, one state is for $x$-polarized wave manipulation, and the other is for $y$-polarized incidence. In addition, the authors in [20] designed a metasurface that can independently manipulate orthogonal linearly polarized terahertz waves by reconfiguring reflection patterns. A series of graphene-strips-based unit cells form the basis of the proposed design. In addition, Zhu et al. proposed a novel design method of aperture-multiplexing metasurfaces using a Back-Propagation Neural Network, which can obtain independent wavefront control of orthogonally polarized electromagnetic waves [21]. For this purpose, they suggested a metasurface based on a modified Jerusalem Cross structure, which decouples orthogonal interactions by boosting the effective inductances of each of the two Jerusalem Cross branches. Due to the reduced orthogonal couplings, the redesigned Jerusalem Cross structure can independently manipulate orthogonally polarized waves. Furthermore, an anisotropic matrix metasurface consisting of asymmetric metal cross particles with simultaneous dual-polarization anomalous reflections is proposed in [22]. There have also been several other studies focused on the development of metasurfaces that are polarization-dependent, such as [23–34]. To the best of our knowledge, there is no designed metasurface in the infrared spectrum capable of manipulating both polarizations simultaneously.

In the past, wireless communication relied chiefly on electromagnetic plane waves [35]. There is also angular momentum in electromagnetic waves, consisting of spin angular momentum (SAM) and orbital angular momentum (OAM). As a wavefront with a spiral phase, the OAM has received a great deal of research attention [36–38]. They can carry different modes (topological charges) independently. Beams with different OAM modes are orthogonal and can be multiplexed/demultiplexed together. As a result, they can increase capacity without relying on traditional resources like time and frequency. In future wireless communication networks, OAM with multiple orthogonal topological charges is expected to bridge a new way to increase spectrum efficiency significantly. Several experiments have recently demonstrated the feasibility of OAM wireless communications [39,40]. According to [41], OAM multiplexing can achieve high capacity in mm-wave communications. Additionally, OAM-based wireless communication research includes mode detection, mode separation, axis estimation and alignment, mode modulation, OAM-beam convergence, etc. [35]. A significant increase in spectrum efficiency can be achieved by combining multiple-input and multiple-output (MIMO) multiplexing with OAM. The spectrum efficiency of wireless communications can be improved by combining MIMO-based spatial multiplexing and OAM multiplexing, as demonstrated in [40,42–44].

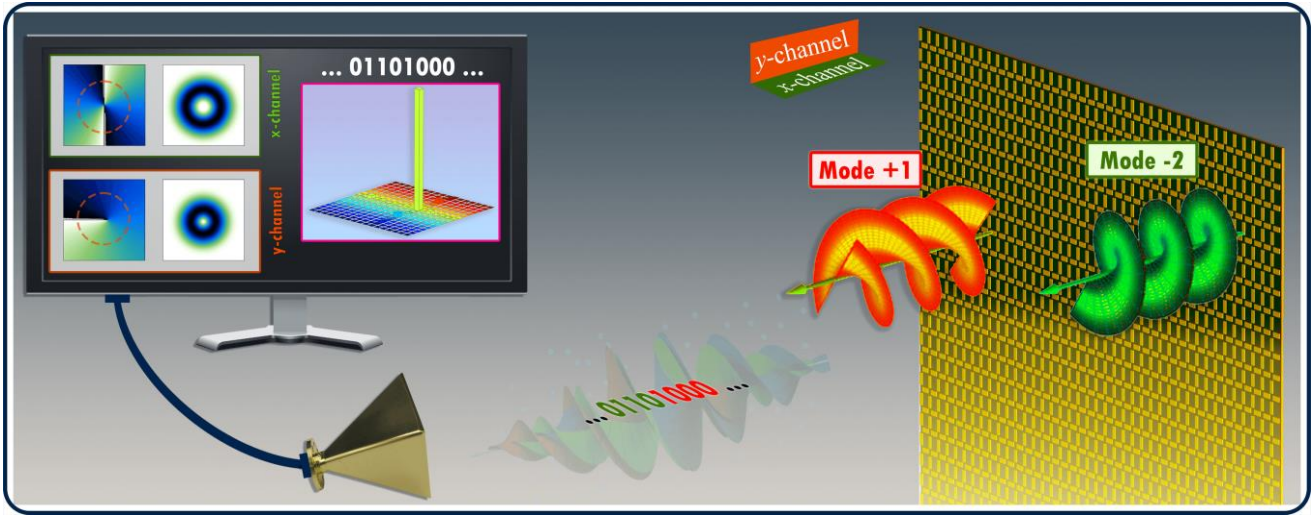

**Fig. 1.** The schematic design of the proposed metasurface for mode-division OAM MIMO communication by using polarization-dependent metasurface based on Indium-Tin-Oxide.

In this article, we survey the architecture of reconfigurable metasurface for imaging applications and wireless nearfield communications. The reflected wave can be controlled separately for each polarization using the polarization-dependent metasurface. We embedded the thin layers of indium-tin-oxide (ITO) wings perpendicular to each other in our design. Hence, the embedded wings in $x$- and $y$-direction can interact only with $x$- and $y$-polarized waves. Furthermore, the corresponding polarization can be controlled by tuning the voltage connected to each wing without affecting cross-polarization. Several examples are presented to illustrate how the design can be applied to lenses. The metasurface mimics the moving lens that can change its focal point without moving physically. Then we demonstrate how MIMO communication in two orthogonal channels can increase the transmission capacity of the communication system, as shown in **Fig. 1**. Our study demonstrates the advantages of OAM-based wireless communications in transmitting information through the channel. Furthermore, we give a novel OAM-modes-based orthogonal multiuser access framework and evaluate the obtained data efficiency in the receivers.

# II. Indium-Tin-Oxide

## 1. ITO Characteristic

Metasurfaces with active tunable properties can be used to extend the applications of devices. For example, spatial light modulators, dynamic beam steering, reconfigurable imaging, and reconfigurable pulse shaping are examples of externally controlling the phase and/or amplitude of the light that the individual antennas on the metasurfaces emit, resulting in a dynamic wavefront transformation of light [45]. Active materials such as transparent conducting oxides [46], graphene [47,48], phase-change materials (e.g., GeSbTe) [49], and liquid crystals [50] are an example of tunable materials. The optical properties of these materials can be switched using external stimuli.

Indium-tin-oxide (ITO), aluminum-zinc-oxide (AZO), gallium-zinc-oxide (GZO), and indium-doped-zinc-oxide (IZO) as transparent conducting oxide materials (TCOs) were investigated to design metamaterials in the near-infrared (NIR) spectral range and also as an option for plasmonic resonances engineering. Among these materials, ITO is the most well-known TCOs deployed in electronic, optoelectronic, and mechanical applications. Utilizing

ITO as an active material with different amounts of carrier density can be useful in tunable devices. In this way, a metasurface with this combination can be made adjustable by externally applying voltage.

Two distinct categories can be considered for the earlier applications of ITO. In the first group, a multilayer metal-oxide-semiconductor (MOS) structure deposited on a silicon waveguide creates plasmonic modulators. In these optical devices, the accumulation layer of ITO controls the plasmonic gap modes. In [51], an ultra-compact "PlasMOStor" made up of an ITO-filled slot plasmonic waveguide has been presented. The second group consists of a sandwiched insulator-ITO layer between two metal electrodes [39]. In metal-insulator-metal (MIM) structures, the confined light experiences a significant phase shift due to Fabry-Perot-like resonance [52,53]. Consequently, these features can be used to control the light-matter interaction. Substituting the insulator with an insulator-ITO layer opens a window of opportunity for generating a reconfigurable antenna with tunable capability. In [54], electrically tunable absorption has been experimentally investigated by depositing a thin layer of ITO over an array of gold nanostrip antennas. It can be inferred from [55] that ITO integration can guide the impinging beam into a MIM structure toward a desired diffracted angle at the selected single frequency by electrically modulating the ITO via the field-effect technique. In this case, both pre-depositional and post-depositional processes and the utilized technique control different structural, electrical, and optical properties of the ITO film (resistivity, transmittance, refractive index, etc.). The most applicable and frequent technique is sputtering, among various methods such as spray pyrolysis, screen printing, and chemical vapor deposition for ITO deposition. All the oxygen content, the tin-to-indium ratio, growth temperature, doping impurity, sputtering power, and pressure help to manage the sputtering process of ITO. Accordingly, different papers have reported the various values for the dielectric function of ITO [54,56–59].

Utilizing a Drude function $\varepsilon_{ITO} = \varepsilon_\infty - \omega_p^2/(\omega^2 + i\omega\Gamma)$ can be helpful to have an accurate optical model for conducting oxide ITO in which background permittivity ($\varepsilon_\infty$), collision frequency ($\Gamma$), and plasma frequency ($\omega_p$) are selected according to the data fitting of the deposited ITO film experimental data. The plasma frequency is related to the carrier density $N$ by $\omega_p^2 = Ne^2/(\varepsilon_0 m^*)$, where $m^* = 0.35m_0$ is the electron's effective mass, $m_0$ indicates the electron rest mass, and $e$ shows the electron charge. Due to the considerable significance of plasma frequency in the NIR regime, a range of $10^{20} - 10^{21}$ cm$^{-3}$ for the carrier concentration should be considered. In this paper, the optical parameters of ITO layers are chosen as $\varepsilon_\infty = 2.17$, and $\Gamma = 190$THz [54,58].

It has been shown in [60,61] that gate voltage bias influences carrier density in the ITO layer based on the Thomas-Fermi screening theory, in which the carrier density change is averaged over the entire ITO layer ($\approx 5\%$ increase in the average carrier density [62]). Another approach is to consider an ultra-thin accumulation layer at the insulator-ITO interface, where the refractive index changes with applied field intensity [59]. The ITO loss can be influenced by various factors, such as the gate-induced injection of carriers into the ITO and the grain size of ITO, which can be controlled by ITO deposition thickness and post-annealing temperature [63,64]. Additionally, it was shown that increasing ITO grain size reduced the resistivity of ITO film [63]. The uniformity of electrical and optical properties of ITO film and the level of sheet resistance is controllable during the fabrication process [64,65].

## 2. Reconfigurable Unit Cell Design

Electromagnetic waves in the terahertz regime can excite prominent plasmonic resonances in ITO, but these wave-ITO interactions need to be further enhanced for practical applications. Therefore, we design a unit cell that greatly enhances wave-ITO interactions by employing the Fabry–Perot resonant principle. The proposed Fabry–Perot resonant-based unit-cell consists of an ITO layer and alumina, which is placed between an optically thick gold substrate (back mirror) and two upper orthogonal gold strips. So, when terahertz waves illuminate the top layer, plasmonic resonances can be excited. In the bottom ground, the metallic material is embedded so that the waves can be completely reflected. The proposed metasurface is constructed by periodically extending the ITO-based unit cells along both $x$- and $y$-directions, as shown in the inset of **Fig. 2(a)**. It's worth mentioning that two gold strips must be arranged perpendicularly to provide independent control of $x$- and $y$-polarized incident waves. Also, two ultra-thin accumulation layers with a thickness of $5 \pm 1$ nm must be considered at the insulator-ITO interface, where the refractive index modifies with the applied field intensity [59]. In this paper, the plasma frequency of the accumulation layer changes between $1.35 \times 10^{15}$ rad/s $3.05 \times 10^{15}$ rad/s, which is realizable based on the presented experimental results in [55,59]. Also, the plasma frequency of the unbiased ITO layer is considered as $1.652 \times 10^{15}$ rad/s. It should be noted that the breakdown voltages for various thicknesses of the Al2O3 layer have been examined in the Supporting information of [66]. In this work, we assume 12 nm-thick alumina, which leads to a breakdown voltage higher than 4 V. Also, the optical behavior of gold is described through its dispersive permittivity, which can be approximated with the Drude model in the NIR spectral region as: $\varepsilon_{r,Au}(\omega) = \varepsilon_\infty - \omega_{p,au}^2(\omega(\omega + i\Gamma))^{-1}$, with $\varepsilon_\infty = 1.53$, the plasma angular frequency is $\omega_{p,au} = 2\pi \times 2.069 \times 10^{15}$ rad/s, and the collision frequency is $\Gamma = 2\pi \times 17.64 \times 10^{12}$ rad/s [67,68]. When the thickness of the back mirror substrate is more than the skin depth of gold in the operating frequencies, it does not affect the structure's optical characteristics. Therefore, it is fixed at 100 nm in the proposed unit cell designs. The permittivity of the Al2O3 is considered a constant value of $\varepsilon_{Al2O3} \approx 3$ in the NIR region.

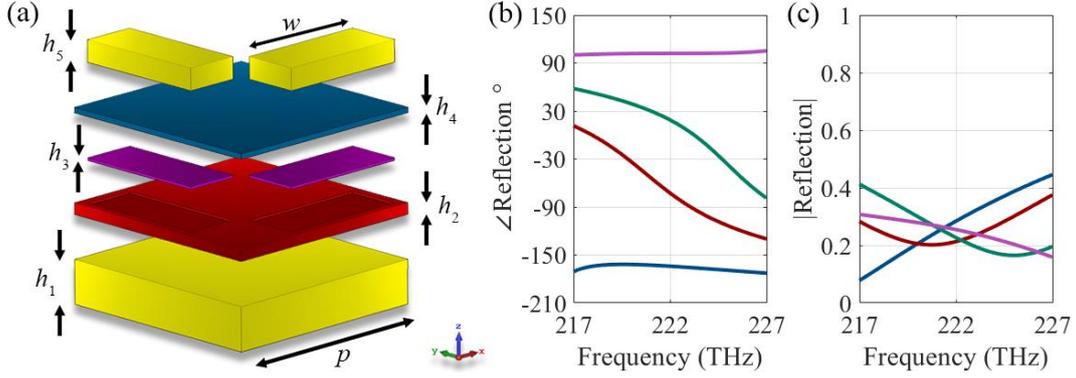

**Fig. 2.** The schematic view of reconfigurable ITO-based unit cell. The (a) phase (b) amplitude spectra of the reflection coefficient of ITO-based unit cell with different plasma frequencies. In these plots, digital states of "00", "01", "10", and"11" are indicated by blue, red, green and purple curves respectively.

While there are many complex structures, we chose two orthogonal ITO structures since they may be the simplest structure that can effectively reduce the mutual coupling of the two orthogonal polarizations. We consider the basic geometry of a single MIM unit cell with the period of $p$, as shown in the inset of **Fig. 2(a)**. The unit cell period shown is p $= 500$ nm, and the wing length is $w = 340$ nm. Also, this figure exhibits the side view of the proposed unit cell, where parameters $h_5 = 50$ nm, $h_4 = 12$ nm, $h_2 = 26$ nm and $h_1 = 100$ nm represent the thickness of gold strips, alumina, ITO, and the bottom metallic ground plane. Concretely, we control the plasma frequency of $\omega_{p,1}$ to control the first strip ($x$-direction) and plasma frequency of $\omega_{p,2}$ to control the second strip ($y$-direction) respectively. Therefore, a numerical simulation based on CST Studio Suite is applied to verify the performance of the proposed design. The phase and amplitude of the infinite array of elements' reflection coefficients are extracted using simulation. The periodic boundary conditions are applied in the $x$- and $y$- directions to incorporate the mutual coupling effect between adjacent cells, and the Floquet ports are also assigned to the $z$-direction.

The different plasma frequencies of the ITO wings that are applied to each unit cell to achieve a wide reflection phase of 270° are $\omega_p = 3.05 \times 10^{15}$ rad/s for the "00" state (blue curve), $\omega_p = 1.35 \times 10^{15}$ rad/s for the "01" state (red curve), $\omega_p = 1.9 \times 10^{15}$ rad/s for "10" state (green curve), and $\omega_p = 2.15 \times 10^{15}$ rad/s for the "11" state (purple), respectively. **Fig. 2(b)** and **(c)** show the simulated reflectivity and reflection phase at 222 THz for both $x$- and $y$-polarized normally incident waves with fixed $\omega_{p,2}$ and varied $\omega_{p,1}$. It demonstrated acceptable $x$-polarized wave reflectivity above 0.2 and a reflection phase span of 270°. Simultaneously, $y$-polarized reflectivity and phases are almost kept constant. It means that tunning $\omega_{p,1}$ will affect the reflection properties of the $x$-polarized wave while keeping $y$-polarized wave unchanged. Similarly, according to the symmetrical structure of the proposed unit cell, $\omega_{p,2}$ will only affect the reflection properties of $y$-polarized waves. Therefore, we conclude that the reflectivity and reflection phases of $x$- and $y$-polarized waves can be manipulated independently by tuning the parameters of $\omega_{p,1}$ and $\omega_{p,2}$, respectively. Importantly, the obtained phase span of 270° is sufficient to provide an exceptional level of performance when used for practical applications. Moreover, the plasmonic resonance of metals becomes less noticeable due to the lower interaction between electrons and waves at terahertz frequencies. As a result, the reflectivity of metallic reflection structures operating at higher terahertz frequencies is frequently less than 30% [38].

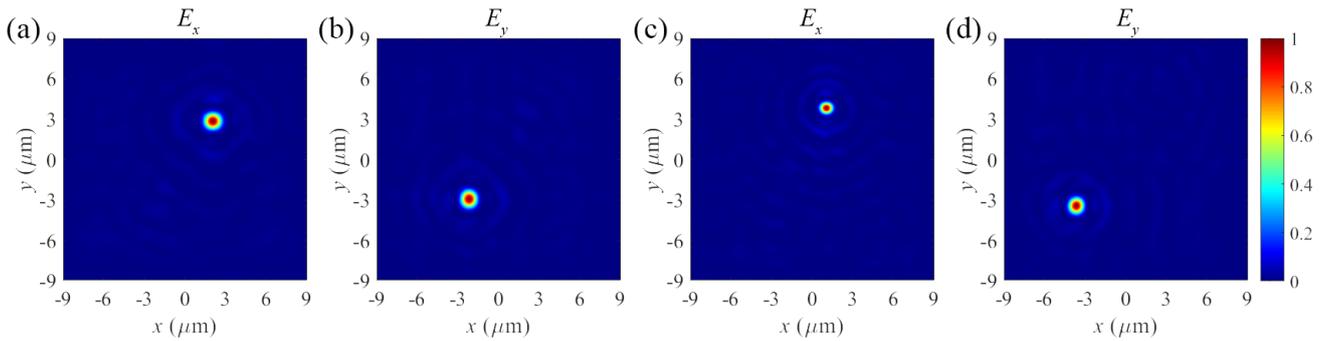

**Fig. 3.** Two-dimensional normalized electric field distributions at the focal planes for (a), (b) the first example at $x_1 = 2$ μm, $y_1 = 3$ μm, $z_1 = 10$ μm and $x_2 = -3$ μm, $y_2 = -3$ μm, $z_2 = 10$ μm and (c), (d) the second example at $x_1 = 4$ μm, $y_1 = 1$ μm, $z_1 = 4$ μm and $x_2 = -3.5$ μm, $y_2 = -3.5$ μm, $z_2 = 8$ μm. (a) and (b) are calculated on the $z = 10$ μm, (c) is calculated for $z = 4$ μm and (d) is calculated for $z = 8$ μm.

## III. Virtually Moving Metalens

### 1. Focus Formulation

Consider a reflective metasurface illuminated by a normally incident plane wave. To make flat metalenses, the waves reflected from the metasurface must interfere at the desired focal point, similar to those transmitted by a conventional lens. At any point on the metasurface, the required phase profile can be written as

$$\Phi(i,j) = k_0 \left( \sqrt{\left(x_i - x_f\right)^2 + \left(y_j - y_f\right)^2 + z_f^2} - z_f \right) \tag{1}$$

where $k_0$ is the incident beam wavenumber in a vacuum, $i$ and $j$ are the element index number on the metasurface, $x_i$ and $y_j$ are the center of the corresponding unit cell, and $x_f$, $y_f$ and $z_f$ indicate the position of the focal point on the focal plane. In this equation, we assume that the metalens is located on the plane of $z = 0$. A metalens phase profile is obtained by rounding the phase value to the closest digital state.

### 2. Arbitrary Dual-pol Concentration

As a result of the Huygens principle, any radiative element can be considered a secondary source. The illuminated wave can be redirected to the desired focal points by irradiating the wave to the metalens. In our simulations, we apply a 45-degree slant incident plane wave as an excitation in terahertz. Slant linear polarization can be seen as the sum of a horizontally polarized linear wave and a vertically polarized linear wave with the same amplitude and phase. The $x$-polarized wave interacts with the horizontal part of the metasurface with the desired reflection phases. At the same time, the $y$-polarized component of the illuminated wave interacts with the vertical wing of the metasurface. The horizontal and vertical illuminations can have separate functions because they are orthogonal.

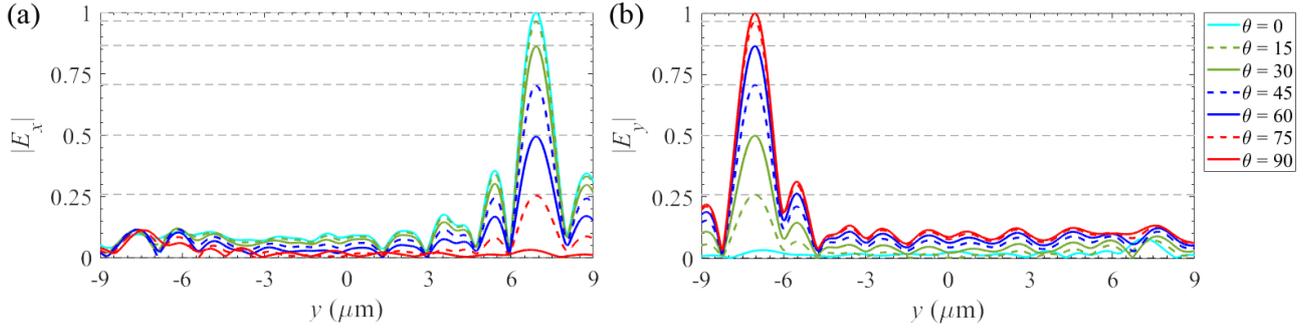

**Fig. 4.** Evaluated electric field on the curve for power modification examples with different incident angles. (a) $x$-polarization, (b) $y$-polarization.

This means that the surface based on the two ITO wings can steer the $x$- and $y$-polarized components of the incident wave with slant linear polarization in two separate directions, respectively.

In order to focus the slant wave on two focal points of $(x_1, y_1, z_1)$ and $(x_2, y_2, z_2)$ simultaneously, we should obtain the metasurface phase profile through **Eq. (1)** in the $x$-direction for the first focal point and in the $y$-direction for the second focal point. For example, we calculated the continuous phase profile required to concentrate the reflected beam at $x_1 = 2$ μm, $y_1 = 3$ μm, and $z_1 = 10$ μm for vertical polarization. We also calculate the continuous phase profile to concentrate the reflected beam at $x_2 = -3$ μm, $y_2 = -3$ μm, and $z_2 = 10$ μm for the horizontal polarization. Then we translate the obtained phase distribution to the corresponding digital states to construct the metalens and simulate the electric field distribution in the desired points. The simulated results for the electric field distribution at the focal plane of $z = 10$ μm are shown in **Fig. 3(a)** and **(b)**. As we can see in these plots, the metasurface produces an efficient electric field concentration very close to the desired focal points. The amount of lateral lobe and the accuracy of the focal point position can be improved by larger metasurface sizes or smaller unit cell dimensions [70].

Alternatively, we arrange the elements to concentrate energy at two different focal points located at two different focal planes. The first focal point for the $x$-polarization is selected at $x_2 = 4$ μm, $y_1 = 1$ μm, and $z_1 = 4$ μm. Also, the second focal point for the $y$-polarization is chosen at $x_2 = -3.5$ μm, $y_1 = -3.5$ μm, and $z_2 = 8$ μm. When determining the focal length, we must be careful that the focal point distance from the metasurface and the position of the focal point in the half-space are important parameters that can affect the available focusing power and beamwidth. Therefore, as we move away from the metasurface, the diameter of the focal region will increase, and the maximum value of the power will decrease. **Fig. 3(c)** and **(d)** show the normalized electric field distribution obtained by the simulation in two focal planes of $z = 4$ μm and $z = 8$ μm, respectively. These examples demonstrate how the focal point can be controlled dynamically and how each polarization can be controlled separately.

## 3. Power Control

In this section, we divide the incident power between two polarizations by varying the incident angle. As an example, the elements are arranged to redirect the $x$-polarized wave into $x_1 = 0$ μm, $y_1 = 7$ μm, $z_1 = 6$ μm and $y$-polarized wave into $x_1 = 0$ μm, $y_1 = -7$ μm, $z_1 = 6$ μm. If we illuminate the wave with the slant angle of $\theta$, the radiant vertical polarization power is obtained with $E_0 \sin \theta$ and the radiant horizontal polarization power is obtained with $E_0 \cos \theta$, where $E_0$ indicates the amplitude of the incident wave. So, the amount of electric field in horizontal and vertical polarization depends on the radiation angle of the illuminated wave. The total normalized field in horizontal and vertical polarization is calculated by $\sqrt{E_0^2 \sin^2 \theta + E_0^2 \cos^2 \theta} = E_0$.

Let us consider a slant wave with $\theta = 60°$. Then we can divide the electric field components to $E_x = E_0 \cos 60° = 0.5 E_0$ for horizontal component and $E_y = E_0 \cos 30° \approx 0.866 E_0$ for the vertical component. The electric field at vertical polarization is approximately 1.732 times larger than the electric field at horizontal polarization at 60 degrees of radiation. In this case, the input power at vertical polarization is greater than the output power at vertical polarization. As a result, the focal point of radiation with horizontal polarization has lower energy than the focal point of vertical polarization.

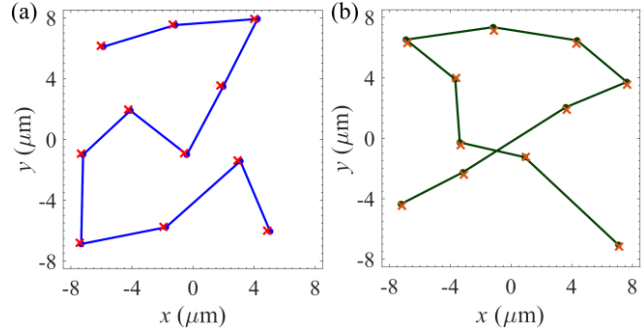

**Fig. 5.** Selected focal points in the focal plane for (a) $x$-polarization and (b) $y$-polarization. In these plots, desired points are represented by '●' and the focal points obtained by the lens are represented by '×'.

According to the given explanation, all the beams emitted at an angle of 0°, 15°, 30°, 45°, 60°, 75°, and 90° degrees to a metasurface can be simultaneously redirected to two separate focal points in two polarizations. The simulated results for these cases are shown in **Fig. 4**. In other words, the amount of power irradiated into horizontal and vertical polarization can be controlled by selecting the proper radiation angle.

## 4. Virtual Moving Metalens

Various methods have been suggested to achieve a lens that can create different focal points. For example, when we want to change the focal point of a fixed lens, we must rotate the lens a bit to reach the new focal point in the new desired position. This mechanical rotation is time-consuming and can waste energy to get a new focal point. Reconfigurable metalenses have an ultrathin thickness and can be tuned dynamically. As a result, attaining reconfigurable functions in a single metasurface has attracted researchers' interest in various terahertz applications. In this section, we utilize a tunable metasurface for manipulating electromagnetic waves and providing wave control without needing mechanical movement. Moreover, the number of focal points and the distance of the focal points with metalens in each polarization can be tuned, which is impossible in conventional lenses. Due to the growing demand for miniaturized and highly integrated systems, this single reconfigurable metasurface with numerous functions is ideal for reconfigurable communication and imaging systems.

In this case, by using the virtual motion of the lens, we can perform parallel imaging in two orthogonal polarizations. For example, we consider a curve by the function of $F(x, y)$, which is discretized by eleven sample dots, represented by the cross sign in **Fig. 5(a)**. In addition, the second curve is defined as $G(x, y)$ that discretized by the same number of points and shown in **Fig. 5(b)**. These two curves are illustrated in **Fig. 5**, which are selected for $x$- and $y$-polarization. We consider the identified points as focal points and calculate the required reflection phase arrangement to obtain desired focal points. Then, we illuminate a radiation wave with slant polarization to the designed metasurface.

As shown in **Fig. 5(a)**, the desired points represented by "●" and the focal points obtained by the metalens represented by "×" are in good agreement. We can say from these results that our lens could act as a virtual-moving metalens array in the $x$-direction. When we illuminate a radiation wave with $y$-polarization to the metalens, we implement the obtained phase arrangements on the vertical strips. As a result, these focal points will be illuminated by the reflected wave from the metasurface, as shown in **Fig. 5(b)**. The designated points of the curve indicated by "●" and the focal points of the metalens with $y$-polarization, indicated by "×", are in good agreement. This section concludes that this structure can be considered a parallel virtual moving lens because the points determined from the desired curve and the focal points obtained by the metalens designed for the $x$- and $y$-polarization are in good agreement.

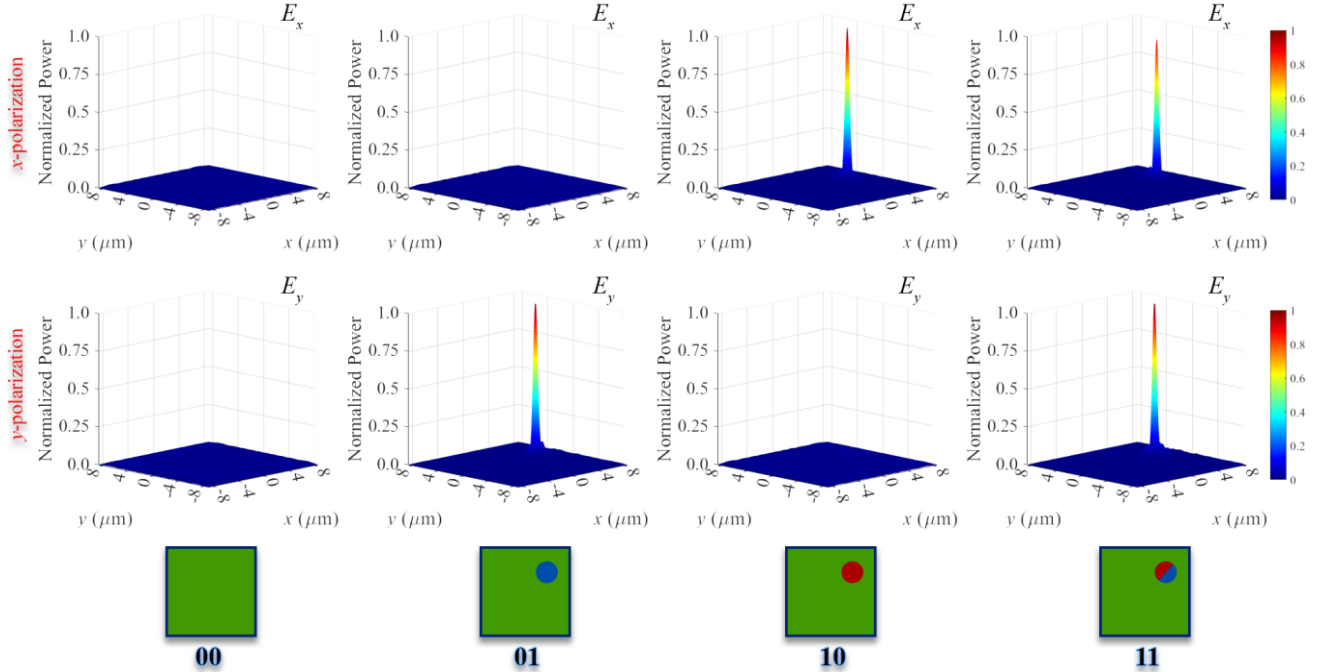

**Fig. 6.** Power distributions in multi-channel communication system. (Column 1) 00, (Column 2) 01, (Column 3) 10, and (Column 4) 11. The selected data point for both polarization is set as $x_1 = 7$ μm, $y_1 = 3$ μm, $z_1 = 8$ μm.

# IV. Near-Field Optical MIMO Communication

## 1. Single Point MIMO Communication

In this section, we use polarization-dependent focal points to build an effective architecture for dual-pol communication, which can be employed widely in MIMO communication. MIMO is an effective method for multiplying the capacity of a communication link using multiple transmission and receiving antennas to exploit multipath propagation. Therefore, MIMO has become an essential component of wireless link standards in the current communication realm [71,72]. In order to transmit information, the focal points must be illuminated by the metalens. For this purpose, by using **Eq. (1)**, we achieve a required phase arrangement to obtain desired focal point in the $x$- and $y$-polarization. As a simple example, the focal points are considered as $x_1 = 7$ μm, $y_1 = 3$ μm, $z_1 = 8$ μm. Let us consider each focal point in each polarization as a data point, which means that we have information transfer with binary code 0 or 1 [73]. We illuminate the terahertz slant wave at a 45-degree angle in order to have equal input power for both polarizations. In order to transmit '00', we do not need to turn on the defined focal points in both polarizations. As a result, the elements arranged in the $x$- and $y$-directions of the metasurface have the same phase, and the wave is returned as a plane wave when it is exposed to this metasurface. We use binary codes to briefly define the phase arrangement of the metasurface in the $x$- and $y$-directions. For example, in this case, the binary code '00' means that the elements are arranged in the same phase in both polarizations. Therefore, no focal points are generated in the focal plane. In the second scenario, we want the information to be transmitted only in the $x$-direction. We arrange the unit cells in the $x$-direction to accomplish this. Thus, the focal point will be turned on only for the $x$-polarization. At the same time, we choose the same phase for the wings in the $y$-direction, so the structure acts as a conductor, and the reflected wave doesn't focus on a certain focal point. Hence, the power does not accumulate at a specific point, which corresponds to the binary code of '01'.

As a third case, we obtain a phase arrangement in the metasurface that merely has a focal point for $y$-polarization. However, there is no energy accumulation in the $x$-polarization. This case is like the binary code of '01', whereas the arrangement of wings in $x$- and $y$-directions is swapped to obtain '10'. In the fourth and last case, the focal point is achieved for both $x$- and $y$-polarization. Thus, power is sent to both focal points in this case, so '11' is transmitted in the channel. We simulate four different arrangements corresponding to four transmitted codes, and the calculated power distribution is illustrated in **Fig. 6**. To detect the transmitted wave in the realistic scenario, we can put a dual-polarization receiver antenna to receive transmitted power in both polarizations.

It is noteworthy that a focal point farther away will have less power [73]. As a result, we must consider more initial power in order to reach the focal point farther away. As we know, the amount of power for $x$- and $y$-polarization can be tuned by varying the incident angle of the slant wave. Thus, if we select data points in two different focal planes, we should choose the optimum slant angle to transmit equal amounts of energy. According to the considered application, different focal points can be defined as data points. For example, we can consider focal points not at the same point or on the same plane for each polarization. Generally speaking, focal points can be defined anywhere with any amount of power. It is possible to transfer complex information with higher capabilities as a result of the ability to define different focal points for $x$- and $y$-polarization. Several cases of information transfer with a single focal point are discussed here, and the rest are beyond the scope of this paper's discussion.

## 2. Near-Field Optical MIMO Communication

Electromagnetic waves transport energy and momentum, and momentum can be classified into two types: linear and angular. One additional component of angular momentum is associated with field polarization, called SAM, and one related to spatial field distribution is called OAM. OAM is one of the fundamental physical properties of electromagnetic waves. It shows the orbital characteristics of electromagnetic rotational degree of freedom and rotation properties of energy. Plane waves essentially generate electromagnetic waves carried by OAM waves with one phase rotation factor $\exp(jm\phi)$, where $m$ is the topological charge (mode) of the OAM beam, and $\phi$ is the azimuthal angle. The azimuthal angle is defined as the angular position on a plane perpendicular to the propagation axis. The wavefront is spiral-shaped as a result of the rotation phase factor. OAM can theoretically be interpreted as a beam with OAM modes that can take not only an integer value but also any non-integer value, and different OAM modes are orthogonal. When the OAM mode is a non-integer, the phase term $\exp(jm\phi)$ can be expressed by the sum of the Fourier series of orthogonal OAM modes. The wavefront phase rotates around the beam propagation direction, and the phase changes $2\pi m$ after a full turn.

Research in OAM in microwave frequencies may focus on multiplexing different modes of OAM at a single frequency to increase the channel capacity of wireless communications. There were controversies surrounding whether OAM multiplexing could work in the far-field and whether it was a subset of MIMO [74]. Despite some controversy, scientists have sought more mature research and development in OAM, OAM antenna, and its system-level implementation in the past decade [74]. Communication through optical fibers, radio communication, free-space optical communication, and acoustic communication can all benefit from OAM [75]. The main applications of this technology are high-speed and high-capacity communications, radar, imaging, and particle manipulation in the optical domain [76].

In contrast to frequency, time, and code-domain orthogonal division, OAM provides a distinct mode-domain to support the orthogonal access of multiple users. OAM allows us to redesign wireless communications since the added orthogonal dimension can improve many aspects of wireless communications. We review three primary advantages of OAM-based wireless communications below [35]. The first advantage is that different OAM modes are orthogonal to each other. Therefore, different OAM modes do not interfere with each other in an ideal scenario. OAM modes can be transmitted in parallel by using polarization-dependent communication. The orthogonality between different OAM modes can enhance spectrum efficiency in wireless communications without requiring more traditional frequency, time, code, and power-domain resources. Additionally, mode-domain resources can be used together with frequency, time, and code-domain resources to significantly increase the spectrum efficiency of wireless communications. As a second benefit, OAM provides an alternative to conventional multiple access methods, such as mode division multiple access (MDMA), without taking up more frequency or time resources. MDMA allows users to access wireless networks orthogonally through different OAM modes. It is expected that future wireless communications will use orthogonal multiple access using mode-domain resources by using this method. Lastly, the conventional frequency hopping techniques for anti-jamming have limitations in a crowded spectrum. The OAM-mode hopping technique has the potential to counter-jamming in the future. In summary, OAM can potentially increase spectrum efficiency, increase the number of users, and improve anti-jamming reliability, but there are still some significant research challenges to address. Current OAM-related research mainly focuses on the line-of-sight scenario, where no fading has been considered. However, there is fading in many practical scenarios, leading to random wavefront phases at the receiver. It is necessary to estimate each mode's phase change when multiple OAM modes experience fading.

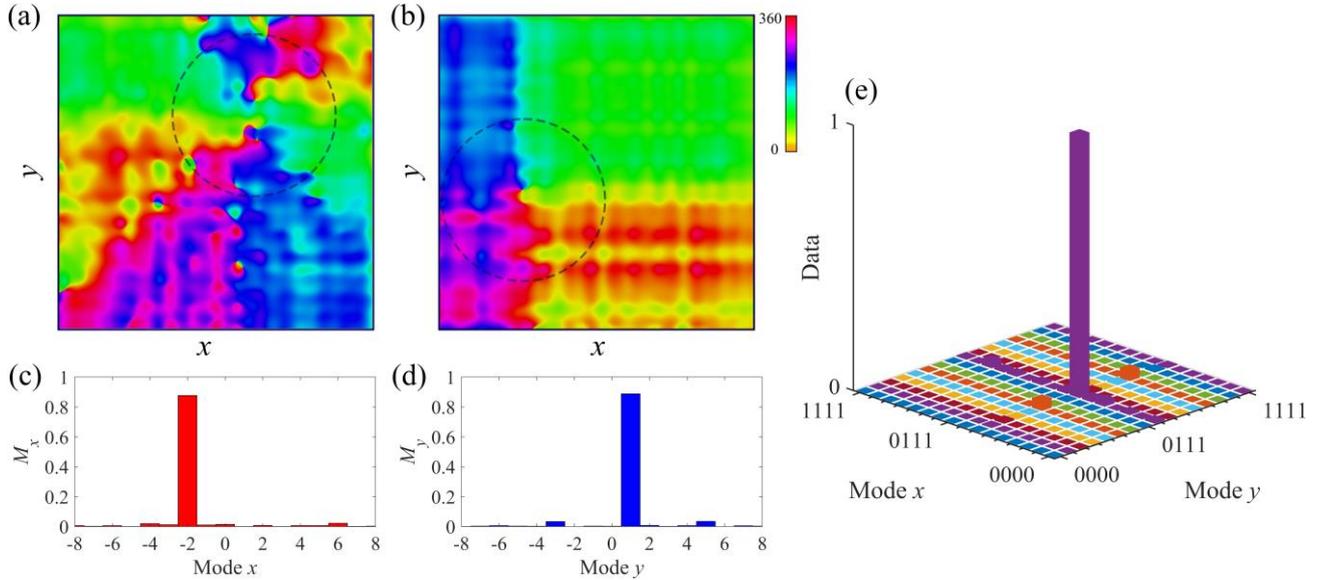

**Fig. 7.** Phase distribution at the receiver plane for (a) $x$-polarization and (b) $y$-polarization. Mode purity diagram corresponding to the orthogonal channel of (c) $x$-polarization and (d) $y$-polarization. (e) Transmitted data through the MIMO channel, i.e., OAM spectrum purity of the generated OAM beams based on the simulated phase distribution.

This section investigates a MIMO communication link that generates OAM beams. Using the proposed method, a capable and secure link based on the orthogonality of OAM beams can be designed for different polarization. In this architecture, we want to transmit 8 bits using the ITO-based metasurface. Thus, 8 bits can be divided into two orthogonal polarizations so that each polarization can carry 4 bits. The transmitted 4 bits can be represented using the OAM beam in which the topological charge is limited between $l = -8$ to $l = 8$, i.e., 16 distinct levels. Let's assume we want to send an octet package of [-2,1] in the communication channel. In this case, the arrangement of elements in the $x$-polarization leads to OAM corresponding to the topological charge of $l = -2$. In the parallel, 4-bit information of $l = 1$ is transmitted in the $y$-polarization. We put receivers at two different locations to demonstrate the method's generality. The first receiver is located at $x = 3.5$ μm and $y = 4.5$ μm to detect $x$-polarization, and the second receiver is located at $x = -6$ μm and $y = -1$ μm to detect $y$-polarization. The simulated electric field in the observation plane of $z = 10$ μm is illustrated in **Figs. 7(a)** and **(b)**. In these plots, the phase distribution of the electric field for $x$- and $y$-polarization are calculated, showing the rotative phase distribution. A spatial analysis using the Fourier transform can be employed to investigate the desired mode's presence accurately. The sampling phase extracted along a circle with a specified radius can be expressed by the Fourier series as

$$\beta(\phi) = \sum_{n=-\infty}^{n=+\infty} a_n e^{jn\phi} \tag{2}$$

$$a_n = \frac{1}{2\pi} \int_0^{2\pi} \beta(\phi) e^{-jn\phi} d\phi \tag{3}$$

where $a_n$ is the coefficient of OAM modes. Then, the OAM spectrum purity can be calculated as follows

$$M = \frac{a_n^2}{\sum_{i=-\infty}^{i=+\infty} a_i^2} \tag{4}$$

Consequently, we calculate and show the OAM spectrum purity of the generated OAM beams based on the simulated phase distribution in **Figs. 7(c)** and **(d)**. According to this bar plot, we recognize the dominant mode and select it as transmitted information. These results demonstrate more than 88% efficiency for OAM spectrum purity, which is excellent for mode detection. Finally, the transmitted data using the proposed architecture is illustrated in **Fig. 7(e)**.

## V. Conclusion

In summary, we proposed an ITO-based metasurface that can control each incident polarization independently. The designed metasurface can be tuned dynamically by varying the biasing voltage. The presented results demonstrate that the delivered power in each polarization changes by varying the incident angle. So, the tunable metasurface can act as a virtually moving reflector controlling transmitted power in each polarization. This reconfigurable metasurface can serve as a virtually moving metalens for parallel imaging. Moreover, we demonstrated the capability of dual-polarization manipulation by using the metalens for MIMO single-point communication in two orthogonal channels. Finally, we provided an alternative to conventional multiple access methods, i.e., mode division multiple access, without taking up more frequency or time resources. By using different modes of OAM, users can access wireless networks orthogonally, increasing bandwidth and capacity. The proposed system has brilliant potential for high-speed MIMO communication.